\documentclass[prd,aps,floats,preprint,preprintnumbers,nofootinbib]{revtex4}
\usepackage{amsmath,amssymb,mathrsfs}
\usepackage{graphicx}
\usepackage{feynmf}
\usepackage{graphics}
\usepackage{amsmath,amscd}
\usepackage{epsfig}
\usepackage{latexsym,oldgerm}
\usepackage{color}
\def\beq{\begin{equation}}
\def\eeq{\end{equation}}
\def\bea{\begin{eqnarray}}
\def\eea{\end{eqnarray}}

\newcommand{\G}{\mathscr{G}}
\newcommand{\Az}{\mathcal{A}_0}
\newcommand{\Ga}{\mathbb{C}\G}
\newcommand{\At}{\mathcal{A}_{\theta}}
\newcommand{\Ft}{F_{\theta}}
\newcommand{\FtMV}{\mathcal{F}_{\theta}^{\mathcal{M},V}}
\newcommand{\FtM}{\mathcal{F}_{\theta}^{\mathcal{M}}}

\newcommand{\Dt}{\Delta_{\theta}}

\newcommand{\AtMV}{\mathcal{A}_{\theta}^{\mathcal{M},V}}
\newcommand{\AtV}{\mathcal{A}_{\theta}^{V}}
\newcommand{\AtM}{\mathcal{A}_{\theta}^{\mathcal{M}}}
\newcommand{\phit}{\phi_{\theta}}
\newcommand{\dx}{{\rm d}}
\newcommand{\Gv}{\mathscr{S}_{\theta}^V}
\newcommand{\e}{{\rm e}}
\newtheorem{de}{Definition}

\begin{document}
\small
\preprint{SU-4252-882 \vspace{1cm}} \setlength{\unitlength}{1mm}
\title{Twisted Quantum Fields on Moyal and Wick-Voros Planes are Inequivalent  
\vspace{0.5cm}}
\author{ A. P.
Balachandran$^{a,b}$}\thanks{C\'atedra de Excelencia\\$\ddagger$bal@phy.syr.edu} \author{M. Martone$^{a,c}$}\thanks{mcmarton@syr.edu}
\affiliation{$^{a}$Department of Physics, Syracuse University, Syracuse, NY
13244-1130, USA\\
$^{b}$Departamento de Matem\'aticas, Universidad Carlos III de Madrid, 28911 Legan\'es, 
Madrid, Spain\\
$^{c}$Dipartimento di Scienze Fisiche, University of Napoli and INFN, Via Cinthia I-80126 Napoli, Italy}
\begin{abstract}
\vspace{0.5cm} The Moyal and Wick-Voros planes $\AtMV$ are $*$-isomorphic. On each of these planes the Poincar\'e group acts as a Hopf algebra symmetry if its coproducts are deformed by twist factors $\Ft^{\mathcal{M},V}$. We show that the $*$-isomorphism T: $\AtM\to\AtV$ does not also map the corresponding twists of the Poincar\'e group algebra. The quantum field theories on these planes with twisted Poincar\'e-Hopf symmetries are thus inequivalent. We explicitly verify this result by showing that a non-trivial dependence on the non-commutative parameter is present for the Wick-Voros plane in a self-energy diagram whereas it is known to be absent on the Moyal plane (in the absence of gauge fields) \cite{bal-statuv-ir, bal,babar}.

Our results differ from these of \cite{Lizzi} because of differences in the treatments of quantum field theories.

\end{abstract}
\maketitle
\section{INTRODUCTION}\label{sec:intro}
General considerations on the coexistence of Einstein's theory of relativity and basic quantum physics, namely Heisenberg's uncertainty principle, strongly suggest that at energy scales close to the Planck scale, spacetime becomes noncommutative \cite{Doplicher}. We can model such
spacetime noncommutativity by the commutation relations
\beq \label{UV1}
[\widehat{x}_{\mu}, \widehat{x}_{\nu}] = i \theta_{\mu \nu} 
\eeq
where $\theta_{\mu \nu} = - \theta_{\nu \mu}$ are constants and 
$\widehat{x}_{\mu}$ are the coordinate functions on $\mathbb{R}^n$ on which the Poincar\'e group acts in a standard manner for $\theta_{\mu\nu}=0$:
\beq
\widehat{x}_{\mu}(x) = x_{\mu}. 
\eeq

There are many noncommutative algebras of functions on spacetime with different products (``$*$-products'') which realize the commutation relations (1). But under suitable conditions, they are all $*$-isomorphic. Two such typical algebras are the Moyal and Wick-Voros planes $\At^{\mathcal{M}}$ and $\At^V$. We recall their $*$-isomorphism in section 2.

Relation (\ref{UV1}) brings with it another problem: at first sight it seems that the noncommutativity of spacetime coordinates violates Poincar\'e invariance. The L.H.S. of (\ref{UV1}), in fact,  transforms in a non-trivial way under the standard action of the Poincar\'e group whereas the R.H.S. does not. But there is still a way to act with the Poincar\'e group on $\At^{\mathcal{M},V}$ if its action is changed to the so-called twisted action \cite{chaichian,wess,sasha}. This modification changes the standard Hopf algebra associated with the Poincar\'e group (the Poincar\'e-Hopf algebra $H\mathscr{P}$) to a twisted Poincar\'e-Hopf algebra $H_{\theta}^{\mathcal{M},V}\mathscr{P}$ ($H_0^{\mathcal{M},V}\mathscr{P}\equiv H\mathscr{P}$). 

In this paper, we establish that the $*$-isomorphism T:$\At^{\mathcal{M}}\to\At^V$, which isomorphically map $H_{\theta}^{\mathcal{M}}\mathscr{P}\to H_{\theta}^V\mathscr{P}$, does not extend to an unitary equivalence of the two Fock spaces. Quantum field theories on these planes with twisted Poincar\'e-Hopf symmetries are thus not equivalent. We explicitly demonstrate this result by showing that whereas a non-trivial dependence on $\theta_{\mu\nu}$ of a self-energy diagram is absent for the Moyal plane in the absence of gauge fields, that is not the case for the Wick-Voros plane.

Our results differ from those of \cite{Lizzi} because of differences in approach to quantum field theories, \cite{Lizzi} following the treatment in \cite{Vitale}.

\section{on Twist-deformations of algebras}

The standard algebra ${\cal A}_{0}\equiv(\mathscr{F}(\mathcal{M}),m_0$) of functions on the Minkowski space-time ${\cal M}\cong \mathbb{R}^{4}$ is commutative with $m_{0}$ being the point-wise multiplication:
\beq
m_{0} (f \otimes g)(x) = f(x)g(x).
\eeq

In order to get relation (\ref{UV1}), we need to deform the product $m_0$ into a noncommutative one. This can be done using the so-called {\it twist deformation} \cite{drinfeld}. Let us call the twist and multiplication map $m_{\theta}^{\mathcal{M},V}$ and $\mathcal{F}_{\theta}^{\mathcal{M},V}$ for $\At^{\mathcal{M},V}$. Then:
\beq\label{UV2}
m_{\theta}^{\mathcal{M},V}\equiv m_0\circ\mathcal{F}_{\theta}^{\mathcal{M},V}
\eeq
Hereafter we will call $\cal{F}^{\cal{M}}_{\theta}$ and $\mathcal{F}_{\theta}^{V}$ respectively the Moyal and Wick-Voros twists and call the Moyal and Wick-Voros algebras $\At^{\mathcal{M}}$ and $\At^{V}$.

In the following, for the sake of simplicity, we will work in $2$-dimensions. The generalization to arbitrary dimensions will be done at the end.  In 2-dimensions, we can write $\theta_{\mu\nu}$ as
\bea\label{UV10}
\theta_{\mu\nu}&=&\theta\epsilon_{\mu\nu}\\
\epsilon_{01}=&-&\epsilon_{10}=1
\eea
where $\theta$ is a constant. Then the two twists assume the form:
\bea\label{UV8}
&&\quad\mathcal{F}^{\mathcal{M}}_{\theta}=\exp\frac{i}{2}\theta[\partial_x\otimes \partial_y-\partial_y\otimes \partial_x]\\\label{UV9}
&&\mathcal{F}^{V}_{\theta}=\mathcal{F}^{\mathcal{M}}_{\theta}\exp\frac{1}{2}\theta[\partial_x\otimes \partial_x+\partial_y\otimes \partial_y].
\eea

Given the above expressions, an explicit form for the noncommutative product of the functions in the two cases follows immediately,
\bea
&&\qquad(f \star_{\mathcal{M}} g)(x)=m_{\theta} (f\otimes g)(x) = m_0\circ\mathcal{F}^{\mathcal{M}}_{\theta}(f \otimes g)(x)\equiv f(x)\textrm{e}^{\frac{i}{2}\theta_{\alpha \beta}\overleftarrow{\partial^{\alpha}}\otimes \overrightarrow{\partial^{\beta}}} g(x)\quad,\\
&&(f \star_V g)(x)=m_{\theta} (f\otimes g)(x) = m_0\circ\mathcal{F}^V_{\theta}(f \otimes g)(x)\equiv f(x)\textrm{e}^{\frac{i}{2}\left(\theta_{\alpha \beta}\overleftarrow{\partial^{\alpha}}\otimes \overrightarrow{\partial^{\beta}} -i\theta\overleftarrow{\partial^{\alpha}}\otimes \overrightarrow{\partial_{\alpha}}\right)}g(x)
\eea
If we let the $*$-product to act on the coordinates functions, we get in both the cases the noncommutative relation of (\ref{UV1}).

Not all the invertible maps from $\mathcal{A}_{\theta}\otimes\mathcal{A}_{\theta}$ onto itself are allowed as twists. Requiring the deformed product $m_{\theta}$ to be associative, we find a certain set of conditions. The two cases considered above fulfill all of them because the differential operators acting on $\mathcal{A}_{\theta}$ commute among themselves. It can be shown that this is a sufficient condition for $m_{\theta}$ to be associative, though it is not a necessary one. In general, the twist has to be a two-cocycle in the Hochschild cohomology, see \cite{Dito} for further discussion.

We end this section with the mathematical definition of the equivalence of two deformations. It proceeds as follows:

 \begin{de}
Two deformations of the algebra of functions $\mathcal{A}_{\theta}\equiv(\mathscr{F} (\mathcal{M}),m_{\theta})$ and $\mathcal{A}'_{\theta}\equiv(\mathscr{F}(\mathcal{M}),m'_{\theta})$ are equivalent if there exists an invertible map {\rm{T}} from $\mathcal{A}_{\theta}\to\mathcal{A}'_{\theta}$ which is compatible with the product in $\mathcal{A}'_{\theta}$, that is:
\beq\label{UV3}
\forall f,g\in \mathscr{F}(\mathcal{M}), \quad {\rm{T}}(f*g)={\rm{T}}(f)*'{\rm{T}}(g)
\eeq
\end{de}
Here $*$ and $*'$ are products induced by $m_{\theta}$ and $m_{\theta}'$.

Writing (\ref{UV3}) in terms of the twist, we get the following condition:
\beq\label{UV7}
m_0\circ\mathcal{F}_{\theta}={\rm{T}}^{-1}\circ m_0\circ\mathcal{F}'_{\theta}\circ({\rm{T}}\otimes{\rm{T}})\quad.
\eeq

For the two deformations described above (namely Moyal and Wick-Voros), such a map T can be found \cite{stern,zachos,fuzzy} explicitly: We have
\beq\label{UV6}
{\rm T}: \At^{\mathcal{M}}\to\At^V,\quad{\rm T}=\exp\left(-\frac{\theta}{4}\nabla^2\right)
\eeq

As a final remark of this section, we want to observe that this map T cannot be seen as a realization of an element of a group since it involves the Laplacian, and that is not an element of its Lie algebra.

Both $\AtM$ and $\AtV$ are $*$-algebras with * being complex conjugation of functions. It is easy to see that T is compatible with this * so that the algebras are $*$-isomporphic.

\section{The Hopf Structure on Symmetry Algebras}

The next step is the specification  of the Hopf algebra structure of a group $\mathscr{G}$ acting on the algebra of functions (see \cite{chari, majid, aschieri} for Hopf algebras). This structure can be determined from considering the action of $\mathscr{G}$ upon the tensor product of functions. This latter is given by the so-called {\it co-product} indicated by $\Delta$. Let $\Ga$ be the group algebra of $\G$. Then $\Delta$ is a homomorphism from $\Ga$ to $\Ga\otimes\Ga$:
\beq
\Delta:\Ga\to\Ga\otimes\Ga\qquad\forall g\in\G,\ f,h\in\mathcal{A}\quad [g]\triangleright (f\otimes h)=(\rho\otimes\rho)(\Delta(g))\circ(f\otimes h)
\eeq

Here $\rho$ is the representation of $\mathbb{C}\mathscr{G}$ on $\mathcal{A}$ and $\triangleright$ indicates the action of the group on functions. 

The explicit form of $\Delta$ is given by asking $\Ga$ to be an {\it automorphism} of the algebra, that is the action of the group has to be compatible with the multiplication rule on $\mathcal{A}$:
\beq\label{UV5}
[g]\triangleright m(f\otimes h)(x)=m([g]\triangleright (f\otimes h))(x)
\eeq

It is easy to see that in the commutative case $\Az$, on group-like elements, the co-product assumes the trivial form: 
\beq
\Delta_0(g)=g\otimes g
\eeq

This co-product can be extended to $\Ga$ by linearity.

The co-product above is not compatible \cite{chaichian,wess,sasha} with the action of $\G$ on the deformed algebra $\At$ . For the twist deformations of the algebra of functions $\AtMV$ considered in (\ref{UV2}), there is a simple ``recipe'' to get deformations $\Dt^{\mathcal{M},V}$ of $\Delta_0$ compatible with $m_{\theta}$:
\beq
\Dt^{\mathcal{M},V}=(F^{\mathcal{M},V}_{\theta})^{-1}\Delta_0 F^{\mathcal{M},V}_{\theta}
\eeq
where $F_{\theta}^{\mathcal{M},V}$ are the realizations of the twist elements $\mathcal{F}_{\theta}^{\mathcal{M},V}$ on $\At^{\mathcal{M},V}$. For the two different twists considered above, namely Moyal and Wick-Voros ones, we will get two different deformations of the group algebra  $\mathbb{C}\G$. Without going deeper into the deformation theory of Hopf algebras, we just note that the deformations we are considering here are very specific ones since we leave invariant the multiplication rule of $\Ga$ deforming only the co-structures of the underlying Hopf algebra (in our case we only change $\Delta_0\to\Dt^{\mathcal{M},V}$ leaving the group multiplication the same).

We can now state when two Hopf algebra deformations of the above type are said to be equivalent \cite{majid}:
\begin{de}

Two twist deformations $F_{\theta}$ and $F_{\theta}'$ are equivalent iff there exists an invertible element $\gamma\in\Ga$ such that\footnote{The condition that follows can be given a cohomological interpretation, namely that the two twists have to be in the same equivalence class. For further details see \cite{majid}.}
\beq\label{UV12}
F_{\theta}=\gamma\otimes\gamma F'_{\theta}\Delta_0(\gamma^{-1})\quad,
\eeq
which for group-like element reduces to
\beq
F_{\theta}=\gamma\otimes\gamma F'_{\theta}\gamma^{-1}\otimes\gamma^{-1}\quad.
\eeq

\end{de}

Note that in terms of co-products,
\beq
\Delta_{\theta}(g)=\gamma\otimes\gamma\Delta'_{\theta}(\gamma^{-1}g\gamma)\gamma^{-1}\otimes\gamma^{-1}
\eeq
which is precisely the condition for the equivalence of the two Hopf algebras $H_{\theta}\G=(\Ga,m,\Delta_{\theta})$ and $H_{\theta}\G'=(\Ga,m,\Delta'_{\theta})$ \cite{majid}

This last equivalence condition is valid even if the co-products do not arise from Drinfiel'd's twists of $\Delta_0=\Delta'_0$

\section{Classical Equivalence of Deformed Algebras with Symmetries}

We can now state the conditions for a (limited) {\it physical equivalence of two deformed algebras with their associated symmetries}. We need not assume that the deformed co-products below arise from Drinfel'd's twists:
\begin{de}
Two deformations, $\At$ and $\At'$, of the algebra of functions, $\Az=\Az'$, which carry the action of the deformed Hopf algebras $H_{\theta}$, $H_{\theta}'$ ($H_0=H'_0$) are \textbf{classically equivalent}\footnote{The reason why we call it classical is because at this level we only deal with functions, that is classical fields. We did not however prove that the actual classical dynamics arising from the two deformations are the same. We hope to discuss this issue elsewhere in more detail in \cite{Marmo}.} if there exists an isomorphism map {\rm T} between the two algebras such that the following condition is fulfilled:
\begin{equation}\label{UV11}
\forall g\in H_{\theta}, \forall f,h\in\At\quad:\quad\mathrm{T}\otimes\mathrm{T}\left(\Delta_{\theta}(g)\triangleright f\otimes h\right)=\Delta'_{\theta}(g_{\mathrm{T}})\triangleright\left(\mathrm{T}(f)\otimes\mathrm{T}(h)\right).
\end{equation}
Here $g\in H_{\theta}$ and ${\rm T}:g\to g_{\rm T}$ is an isomorphism on $\Ga$. It can be written in a diagrammatic form using the following commutative diagram:
\begin{equation}
\begin{CD}
\At\otimes\At @>\mathrm{T}\otimes\mathrm{T}>> \At'\otimes\At'\\
@VV\Delta_{\theta}(g)V @VV\Delta'_{\theta}(g_{\mathrm{T}})V\\
\At\otimes\At @>\mathrm{T}\otimes\mathrm{T}>> \At'\otimes\At'
\end{CD}
\end{equation} 
Here $g\in H_{\theta}$.
\end{de}

In the case of the Drinfel'd twists discussed above, ${\rm T}=\gamma$ and $g_{{\rm T}}=\gamma^{-1}g\gamma$.

The necessity for the characterization of the above equivalence condition as classical\normalcolor\  merits illustration by an example where two classically equivalent deformations, and even that with qualification, lead to differing physics. We now give such an example.

\section{Inequivalence of Moyal and Wick-Voros QFT's}

We first show that at the classical level\ $\AtMV$ are equivalent when Poincar\'e invariance is implemented by twisting the co-products by $F_{\theta}^{\mathcal{M},V}$. This result follows from direct computation of (\ref{UV12}) since\footnote{It is possible to prove, using a non-trivial construction, that the above equivalence is directly implied since the two spacetime algebras are isomorphic with the isomorphism T$\in H$. See \cite{Marmo} for a discussion.} 
\beq
\Delta_0({\rm T^{-1}})=\exp\left(\frac{\theta}{4}\Delta_0(\nabla^2)\right)={\rm T^{-1}}\otimes {\rm T^{-1}}\exp\left(\frac{\theta}{4}2\partial_\mu\otimes\partial^\mu\right)
\eeq
for T as in (\ref{UV6}). Since $[\nabla^2,\partial_{\mu}]=0$ then in the case of Moyal and Wick-Voros:
\beq
({\rm T}\otimes{\rm T})\mathcal{F}_{\theta}^{\mathcal{M}}\Delta_0({\rm T}^{-1})=\mathcal{F}^{\mathcal{M}}_{\theta}\exp\left(-\frac{\theta}{4}2\partial_\mu\otimes\partial^\mu\right)=\mathcal{F}_{\theta}^V
\eeq

To proceed further, let us first show that the $\mathcal{R}$ matrices in the case of Moyal and Wick-Voros planes are exactly the same. Hence the twisted statistics operators are equal for the Moyal and Wick-Voros planes. The proof is as follows. Recalling (\ref{UV8}-\ref{UV9}), the two twists can be written as:
\begin{eqnarray}
&&\qquad\ \mathcal{F}^{\mathcal{M}}_{\theta}=\exp\left(\frac{i}{2}\partial^{\mu}\theta_{\mu\nu}\otimes \partial^{\nu}\right)\\\label{Sta4}
&&\mathcal{F}^{V}_{\theta}=\mathcal{F}^{\mathcal{M}}_{\theta}\exp\left(\frac{\theta}{2}\partial_{\mu}\otimes \partial^{\mu}\right)\equiv\mathcal{F}^{\mathcal{M}}_{\theta}\mathscr{S}_{\theta}^V\label{Sta7}\quad .
\end{eqnarray}

A simple argument \cite{mangano} then shows that the twisted flip operators defining statistics are:
\beq
\tau_{\theta}^{\mathcal{M},V}=(\FtMV)^{-1}\tau_0\FtMV
\eeq

It is then easy to show that the symmetric exponential $\mathscr{S}_{\theta}^V$ multiplying $\FtM$ in (\ref{Sta7}) which differentiates Moyal from Wick-Voros disappears from the $\tau_{\theta}^{\mathcal{M},V}$. Hence $\tau_{\theta}^{\mathcal{M}}=\tau^V_{\theta}$.

It follows then that the creation and annihilation operators in the two cases are twisted (essentially) in the same way\footnote{This remark merits further considerations. A full treatment of quantum fields on $\At^{V}$ will be given in \cite{patrizia}.}. If $\alpha_p$, $\alpha^{\dagger}_p$ and $c_p$, $c^{\dagger}_p$ are the ``dressed'' and ``undressed'' annihilation and creation operators in the Voros case, then\footnote{Upto equivalence \cite{patrizia} which does not affect what follows.}
\begin{equation}\label{Sta5}
\alpha_p=c_p\exp\left({\frac{i}{2}p_{\mu}\theta^{\mu\nu}P_{\nu}}\right)\quad{\rm and}\quad \alpha^{\dagger}_p=\exp\left(-{\frac{i}{2}p_{\mu}\theta^{\mu\nu}P_{\nu}}\right)c_p^{\dagger}
\end{equation}
where $P_{\mu}$ is the total four-momentum operator. For a free field of mass $m$, it is
\begin{equation}\label{Sta10}
P_{\nu}=\int\dx\mu_p p_{\nu}\alpha^{\dagger}_p\alpha_p, \qquad \dx\mu_p=\frac{\dx p}{2|p_0|}, \quad |p_0|=\sqrt{p^2+m^2}.
\end{equation}

We now explicitly check that the $\theta$ dependence in the Wick-Voros case produces non-trivial effects in quantum field theory even when there are no gauge fields. 

Let us first consider the Moyal case. The twisted free field $\phi^{\mathcal{M}}_{\theta}$ of mass $m$ has the mode expansion \cite{patrizia}
\begin{equation}
\phit(x)=\int \dx\mu_p\Big[a_pe_p(x)+a^{\dagger}_pe_{-p}(x)\Big]
\end{equation}
where $a_p$, $a^{\dagger}_p$ are (upto equivalence \cite{patrizia}) $\alpha_p$, $\alpha^{\dagger}_p$ and where $e_p(x)\equiv{\rm e}^{ip\cdot x}$. These fields at the same $x$ must be multiplied with the Moyal $*$ product $*_{\mathcal{M}}$. As a consequence \cite{Vaidya}
\bea
\phit^{\mathcal{M}}*_{\mathcal{M}}\phit^{\mathcal{M}}*_{\mathcal{M}}...*_{\mathcal{M}}\phit^{\mathcal{M}}=(\phi_0\phi_0\cdots\phi_0)\exp\left(\frac{i}{2}\overleftarrow{\partial_{\alpha}}\theta^{\alpha\beta}P_{\beta}\right),\qquad \phi_0=\phi_0^{\mathcal{M}}\quad.
\eea

Using this identity, it has been proved elsewhere \cite{babar,bal-statuv-ir, bal} that in the Moyal case the $\mathcal{S}$-matrix $\mathcal{S}_{\theta}^{\mathcal{M}}$ in the absence of gauge fields is independent of $\theta_{\mu\nu}$.

But this identity fails in the Wick-Voros case because of the extra twist $\Gv$ in (\ref{Sta7}). One finds for a Wick-Voros field $\phi^V_{\theta}$ of mass $m$, 
\bea\label{Sta8}
\phi^V_{\theta}*_V\phi^V_{\theta}=m_0(\phi_0\Gv\phi_0)\exp\left(\frac{i}{2}\overleftarrow{\partial_{\alpha}}\theta^{\alpha\beta}P_{\beta}\right)&=&\phi_0\exp{\left(\frac{\theta}{2}\overleftarrow{\partial}\cdot\overrightarrow{\partial}\right)}\phi_0\exp\left(\frac{i}{2}\overleftarrow{\partial_{\alpha}}\theta^{\alpha\beta}P_{\beta}\right)\\\nonumber \quad \phi_0&\equiv&\phi_0^V
\eea
a formula which can be generalized to the product of $n$ fields:
\beq\label{Sta6}
\phit^V*_V\phit^V*_V...*_V\phit^V=\phi_0\exp{\left(\frac{\theta}{2}\overleftarrow{\partial}\cdot\overrightarrow{\partial}\right)}\phi_0\exp{\left(\frac{\theta}{2}\overleftarrow{\partial}\cdot\overrightarrow{\partial}\right)}...\exp{\left(\frac{\theta}{2}\overleftarrow{\partial}\cdot\overrightarrow{\partial}\right)}\phi_0\exp\left(\frac{i}{2}\overleftarrow{\partial_{\alpha}}\theta^{\alpha\beta}P_{\beta}\right)
\eeq

Here the $\overleftarrow{\partial}$'s differentiate all fields to the left and $\overrightarrow{\partial}$'s differentiate all fields to the right.

We can now evaluate the one-loop contribution to the propagator with interaction $\frac{\lambda}{4!}\phi_{\theta}^V*_V\phi_{\theta}^V*_V\phi_{\theta}^V*_V\phi_{\theta}^V$. Let us call $p$ and $q$ the momenta of the incoming ant outgoing particle, then its $\mathcal{O}(\lambda)$ correction is:
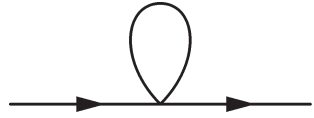
\vspace{2 cm} 

\begin{center}
\begin{fmffile}{selfene}
\begin{fmfgraph*}(40,25)
\fmfpen{thin}
\fmfleft{i1}
\fmf{fermion,label={\it p}}{i1,v1}
\fmf{fermion,label={\it q}}{v1,o1}
\fmf{plain,label={\it k}}{v1,v1}
\fmfright{o1}
\end{fmfgraph*}
\end{fmffile}
\end{center}

In the case of Moyal, this self-energy does not show any $\theta$ dependence. Here instead that is not the case. To see this, we note that
\beq
\int\dx^2x\ \phit^V*_V\cdots\phit^V=\delta^2(p_1+p_2+p_3+p_4)\tilde{\phi}_0(p_1)\cdots\tilde{\phi}_0(p_4)\prod_{a<b}{\e}^{\frac{\theta}{2}p_ap_b}
\eeq
where
\beq
\phi_0(x)=\int\dx^2k\tilde{\phi}_0(k)\e^{ik\cdot x}
\eeq

Hence the last factor in (\ref{Sta8}) becomes 1 and the matrix element $\langle p|i\int\dx^2x\Big(\phit^V*_V\cdots\phit^V\Big)|q\rangle$ has the typical term
\beq
constant\times\delta^2(p-q)\int\frac{\dx^2k}{(2\pi)^2}\frac{1}{k^2+m^2}\left[\prod_{a<b}\exp{\left(\frac{\theta}{2}\vec{p}_a\cdot\vec{p}_b\right)}\right]
\eeq
where one $p_a$ is $p$, another is $-q$ and the remaining $p_c$'s are $k$. So we can see that the terms which mix the external momenta with the internal one cancel each other since they occur in an even number of terms with alternate signs. 

Finally we see that this diagram is proportional to
\beq\label{Sta9}
\delta^2(p-q)\Pi^{(1)}={\rm e}^{-\frac{\theta}{2}p^2}\int \frac{\dx^2k}{(2\pi)^2}\frac{{\rm e}^{-\frac{\theta}{2}k^2}}{k^2+m^2}=\e^{-\frac{\theta}{2}p^2}\int_0^{\infty}\frac{\dx\alpha}{(4\pi)}\frac{{\rm e}^{-\alpha m^2}}{\frac{\theta}{2}+\alpha}
\eeq
where $\Pi^{(1)}$ has been evaluated using the Schwinger parametrization and is finite, since the ``infrared'' divergence has been removed by $\theta$:
\beq
\delta^2(p-q)\Pi^{(1)}=\frac{{\rm e}^{\frac{\theta}{2}(m^2-p^2)}}{4\pi}\int_{\frac{\theta}{2}}^{\infty}\dx\beta\frac{{\rm e}^{-m^2\beta}}{\beta}
\eeq
which gets its usual divergent behaviour in the limit $\theta_{\mu\nu}\to0$ \cite{Minwalla,BaPQ}.

Equation (\ref{Sta9}) shows a dependence on the sign of the symmetric part $\Gv$ since a minus sign in (\ref{Sta4}) will make the diagram even more divergent in Euclidean space than in the standard commutative theory. An explanation of that can perhaps be found in the fact that the Wick-Voros product is obtained from the noncommutative algebra of operators by taking the expectation value in coherent states. The sign in (\ref{Sta4}) is obtained using the normal ordering of operators. The opposite choice, namely the anti-normal ordering, instead gives rise to the minus sign. The divergence of the diagram may then traced to the lack of proper definition of the latter\footnote{We gratefully acknowledge Giuseppe Marmo for having pointed this out.}.

\section{Final Remarks}

We can now briefly outline how to generalize our considerations on the Wick-Voros twist (\ref{UV9}) to $2N$-dimensions\footnote{In 2$N+1$-dimensions, we can always choose $2N+1$ so that $\theta_{\mu,2N+1}$=$\theta_{2N+1,\mu}$=0)}. We can always choose $\hat{x}_{\mu}$ so that $\theta_{\mu\nu}$, now a $2N\times 2N$ antisymmetric matrix, becomes a direct sum of $N$ $2\times 2$ ones. By antisymmetry, these $2\times2$ matrices are of the form (\ref{UV10}) but different 2$\times$2 matrices may have different $\theta$'s. For every such $2\times2$ block, we have a pair of $\hat{x}$'s which can be treated as in the 2-dimensional case above. (Of course there is no twist in any block with a vanishing $\theta$.)

In conclusion when two algebras $\mathcal{A}$ and $\mathcal{A}'$ and its Hopf symmetries are both twisted, the equivalence at the level of classical fields\ using the two deformed structures requires the equivalence of both the twisted algebras and their Hopf symmetries in a precise sense. We have derived this condition. We have then shown that going to quantum theory, as often happens, introduces more complication. We have proven that, although the classical fields on Moyal and Wick-Voros planes satisfy the condition of equivalence given in the paper, their quantum field theories are not equivalent. We have also explicitly shown this inequivalence by calculating a simple Feynman diagram.

\section{Acknowledgements}

We gratefully acknowledge discussions with Fedele Lizzi, Alexander Pinzul, Amilcar R. Queiroz and  Patrizia Vitale. We want also to thank Anosh Joseph and Pramod Padmanabham for encouraging discussions. It is also a pleasure to thank Alvaro Ferraz and the Centro Internacional de F\'isica da Mat\'eria Condensada of Brasilia where this work was started and Alberto Ibort and the Universidad Carlos III de Madrid for their wonderful hospitality and support.

A.P.B. thanks T. R. Govindarajan and the Institute of Mathematical Sciences, Chennai for very friendly hospitality as well.

The work was supported in part by DOE under the grant number DE-FG02-85ER40231 and by the Department of Science and Technology, India.

\bibliographystyle{apsrmp}

\begin{thebibliography}{99}

\bibitem{babar} A. P. Balachandran, A. Pinzul and B. A. Quereshi, {\it UV-IR Mixing in Non-Commutative Plane} Phys, Lett. B, {\bf 634} 4 434-436 (2006) [arXiv:hep-th/0508151]

\bibitem{bal-statuv-ir} A. P. Balachandran, T. R. Govindarajan, G. Mangano, A. Pinzul, B. A. Qureshi, S. Vaidya, {\it Statistics and UV-IR mixing with twisted Poincare invariance}, Phys. Rev.  D {\bf 75} 045009 (2007), [arXiv:hep-th/0608179].

\bibitem{bal} A. P. Balachandran, A. Pinzul and S. Vaidya, {\it  Spin and statistics
on the Groenewold-Moyal plane: Pauli-forbidden levels and transitions}, Int. J. Mod.
Phys. A {\bf 21} 3111 (2006) [arXiv:hep-th/0508002].

\bibitem{Lizzi} S. Galluccio, F. Lizzi and P. Vitale, {\it Twisted noncommutative field theory with the Wick-Voros and Moyal products}, Phys. Rev. D {\bf 78}, 085007 (2008), [arXiv:hep-th/0810.2095].

\bibitem{Doplicher} S. Doplicher, K. Fredenhagen and J. E. Roberts, {\it Spacetime quantization induced by classical gravity}, Phys. Lett. B {\bf 331}, 33-44 (1994).

\bibitem{chaichian} M. Chaichian, P. Kulish, K. Nishijima, A. Tureanu, {\it On a Lorentz-invariant interpretation of noncommutative space-Time and its implications on noncommutative QFT}, Phys. Lett. B {\bf 604}, 98-102 (2004), [arXiv:hep-th/0408069].

\bibitem{wess} J. Wess, {\it Deformed Coordinates Spaces; Derivatives} (2004), [arXiv:hep-th/0408080]

\bibitem{sasha} A. P. Balachandran, A. Pinzul and B. A. Quereshi, {\it Twisted Poincar\'e invariant quantum field theories}, Phys. Rev. D {\bf 77}, 025021(2008) [arXiv:hep-th/0708.1779]. 

\bibitem{Vitale} P. Aschieri, F. Lizzi and P. Vitale, {\it Twisting all the way: From Classical Mechaincs to Quantum Fields}, Phys. Rev. D {\bf 77}, 025037 (2008) [arXiv:hep-th/0708.3002]


\bibitem{drinfeld} V. G. Drinfel'd, Leningrad Math. J. {\bf 1} 321 (1990).

\bibitem{Dito} G. Dito and D. Sternheimer, {\it Deformation quantization: genesis, developments and metamorphoses}, in {\it Deformation quantization (G. Halbout, ed.)}, IRMA Lectures in Math. Physics {\bf I}, 9-54, (Walter de Gruyter, Berlin 2002). math.QA/02201168 (2002)

\bibitem{stern} G. Alexanian, A. Pinzul and A. Stern, {\it Generalized coherent state approach to star products and applications to the Fuzzy Sphere}, Nucl. Phys. B {\bf600} 531 (2001) [arXiv:hep-th/0010187].

\bibitem{zachos} C. K. Zachos, {\it Geometrical evaluation of star products},  J. Math. Phys. {\bf41} 5129 (2000) [arXiv:hep-th/9912238].

\bibitem{fuzzy} A. P. Balachandran, S. K\"urkc\"u\c co\v glu and S. Vaidya, {\it Lectures on Fuzzy and Fuzzy SUSY physics}, World Scientific Publishing, Singapore (2007), \S 3.

\bibitem{chari} V. Chari and A. Pressley, {\it A guide to quantum groups}, Cambridge University Press, Cambridge (1994).

\bibitem{majid} S. Majid, {\it Foundations of Quantum Group Theory}, Cambridge University Press, Cambridge (1995)

\bibitem{aschieri} P. Aschieri, {\it Lectures on Hopf Algebras, Quantum Groups and Twists}, (2007) [arXiv:hep-th/070313v1]

\bibitem{Marmo} A. P. Balachandran, A. Ibort, G. Marmo, M. Martone, in preparation.

\bibitem{mangano} A. P. Balachandran, T. R. Govindarajan, G. Mangano, A. Pinzul, B. A. Quereshi and S. Vaidya {\it Statistics and UV-IR Mixing with Twisted Poincar\'e Invariance}, Phys. Rev. D {\bf 75}, 045009 (2007) [arXiv:hep-th/0608179]

\bibitem{patrizia} In preparation.


\bibitem{Vaidya} A. P. Balachandran, A. Pinzul, B. A. Quereshi and S. Vaidya, {\it Twisted gauge and gravity theories on the Groenewold-Moyal plane}, [arXiv:hep-th/0708.0069]

\bibitem{inv} A.P. Balachandran, B. A. Quereshi, {\it Statistics and Twisted Poincar\'e Invariance in Quantum Theories}, AIP Conf. Proc. {\bf 939}, 305-315, (2007)

\bibitem{Minwalla} S. Minwalla, M. Van Raamsdonk and N. Seiberg, {\it Noncommutative Perturbative Dynamics}, JHEP {\bf 02} 020, (2000), [arXiv:hep-th/9912072]

\bibitem{BaPQ} A. P. Balachandran, A. Pinzul, A. R. Queiroz {\it Twisted Poincar\'e Invariance, Noncommutative Gauge Theories and UV-IR Mixing}, Phys. Lett. B {\bf 668}, 241-245 (2008), [arXiv:hep-th/0804.3588] 





\end{thebibliography}

\end{document}